\documentclass[aps,prd,reprint,longbibliography,nofootinbib,superscriptaddress]{revtex4-2}
\usepackage{graphicx}
\usepackage{bm}
\usepackage{multirow}
\usepackage{array}
\usepackage{amsmath}
\usepackage{mathrsfs}
\usepackage{amssymb}
\usepackage{xcolor}
\usepackage{enumerate}
\usepackage{slashed}
\usepackage{hyperref}
\usepackage{url}
\usepackage{mathtools}
\usepackage{booktabs}
\usepackage{BOONDOX-cal}
\usepackage{orcidlink}
\usepackage{makecell}
\usepackage{cancel}
\allowdisplaybreaks[4]
\hypersetup{colorlinks=true, linkcolor=blue, citecolor=blue, urlcolor=blue}

\begin{document}
	
\title{Spectra and nonleptonic decays of multiply heavy baryons in a finite-size quark–diquark model}  
	
	\author{Yongtao Yu}
	\affiliation{College of Physics, Chengdu University of Technology, Chengdu 610059, China}
	
	\author{Tao Guo}
	\email[Corresponding author: ]{guot17@tsinghua.org.cn}
	\affiliation{College of Physics, Chengdu University of Technology, Chengdu 610059, China}
	
\date{\today}
	

\begin{abstract}
	We investigate the mass spectra and two-body nonleptonic weak decays of ground-state doubly and triply heavy baryons in a nonrelativistic quark-diquark model. To account for the finite size of the diquark, we generalize the convoluted quark-diquark interaction to unequal-mass diquarks using mass-weighted constituent coordinates and distinct pairwise spin-spin couplings. The resulting Schr{\"o}dinger equations provide the baryon masses and wave functions used consistently in the decay calculation. The predicted masses differ from existing lattice QCD results by less than $1.5\%$, and our calculated mass of $\Omega_{cc}^{+}$ is consistent with recent LHCb observations. The calculated hyperfine splittings are systematically slightly smaller than the lattice values, indicating a residual uncertainty in the spin-dependent interaction. Within the factorization approximation, we evaluate the $c\to s$ and $b\to c$ transitions with pion or kaon emission. At leading nonrelativistic order, the vector contribution vanishes for the $3/2^+\to1/2^+$ transitions, while both vector and axial contributions enter the $3/2^+\to3/2^+$ channels. The $b\to c$ spatial overlaps are substantially smaller than their $c\to s$ counterparts. Together with the Cabibbo-Kobayashi-Maskawa hierarchy, this reduces the corresponding $b \to c$ width by approximately two orders of magnitude. In the $3/2^+\to1/2^+$ charm transitions, the leading $|\bm{k}|^3$ dependence enhances the suppression of kaon emission. In bottom transitions with a larger energy release, the smaller relative phase-space difference brings the ratios close to the scaling determined by the Cabibbo-Kobayashi-Maskawa elements and meson decay constants. These results provide benchmark estimates for future studies of multiply heavy baryons and quantify the role of recoil dynamics within the finite-size quark-diquark framework.
\end{abstract}


\maketitle

\section{Introduction}

Heavy-flavor baryons provide a setting in which weak transitions can be studied together with nonperturbative quantum chromodynamics (QCD). 
Systems containing two or three heavy quarks, $QQq$ ($Q=c,b$ and $q=u,d,s$) and $QQQ$, are particularly useful because the heavy-quark masses introduce controlled dynamical scales while the light or strange degrees of freedom remain sensitive to confinement. 
Their small production rates and limited experimental information, however, leave their spectra and decay properties less well established than those of light and singly heavy baryons~\cite{ParticleDataGroup:2024cfk}.

The observation of the doubly charmed baryon $\Xi_{cc}^{++}$ by LHCb in 2017~\cite{LHCb:2017iph}, followed by measurements of additional decay modes~\cite{LHCb:2018pcs,LHCb:2022rpd} and its lifetime~\cite{LHCb:2018zpl}, has motivated extensive studies of doubly heavy baryons~\cite{Hu:2019bqj,Yu:2019lxw,Han:2021gkl,Qin:2021zqx,Guo:2021yws,Cheng:2021qpd}. 
More recently, the LHCb Collaboration reported the first observation of the weakly decaying doubly charmed baryon $\Omega_{cc}^{+}$ in the $\Omega_c^{0}\pi^{+}$ channel with a global significance of $8.7\sigma$ and measured its mass to be $3725.9\pm1.0\pm0.2\pm0.4\pm0.6~\mathrm{MeV}$~\cite{LHCb:2026lnj}.
To date, triply heavy baryons have not yet been experimentally observed, and relevant studies have merely provided complementary probes into heavy-quark dynamics and confinement~\cite{Brambilla:2005yk,Jia:2006gw,Zhang:2009re,Flynn:2011gf,Wei:2015gsa,Yang:2019lsg}.
Baryons of the form $QQs$ connect these systems to a regime in which the third quark constituent is lighter and relativistic effects may be more important.

The spectra of doubly and triply heavy baryons have been investigated with lattice QCD~\cite{Briceno:2012wt,Padmanath:2013zfa,PACS-CS:2013vie,Brown:2014ena} and a variety of phenomenological models~\cite{Ebert:2002ig,Faustov:2021qqf,Ortiz-Pacheco:2023kjn,Yu:2025gdg,Liu:2025jyn,Zhou:2025fpp}. 
Quantitative calculations of their nonleptonic weak decays are less developed. 
Light-front quark model (LFQM)~\cite{Wang:2022,Lu:2024,Zhao:2022vfr} and QCD sum rules~\cite{Wang:2011ae,Najjar:2024ngm} have focused mainly on $c\to s$ transitions of fully charmed baryons, whereas dynamical predictions for $b\to c$ modes such as $\Omega_{bbb}^{-}\to\Omega_{bbc}^{0}\pi^{-}$ remain limited. 
Flavor-$SU(3)$ analyses provide amplitude relations~\cite{Wang:2018utj,Huang:2021jxt}, but absolute widths require additional information about the hadronic matrix elements.

Two related issues are important for these decays. 
First, the large energy release in $b\to c$ transitions produces substantial recoil, making the amplitude sensitive to cancellations in the overlap of the initial and final baryon wave functions. 
Second, many quark-diquark implementations treat the diquark as pointlike~\cite{Lichtenberg:1968zz,Maji:2016yqo,Watanabe:2021nwe}. 
Its finite spatial extent can modify both the effective quark-diquark interaction and the wave functions entering the recoil overlap. 
A consistent treatment of spectroscopy and weak decay therefore requires a framework in which the diquark structure is retained in both calculations.

In this work, we study the spectra and two-body nonleptonic weak decays of ground-state $QQQ$ and $QQs$ baryons in a nonrelativistic constituent quark-diquark model. We use the finite-size convolution introduced in Ref.~\cite{Tourbez:2025hlh} and extend it to unequal-mass diquarks by retaining the mass-weighted constituent coordinates and the distinct pairwise spin-spin interactions. 
This extension is the principal methodological development of the present calculation. 
We solve the radial Schr{\"o}dinger equations with the Numerov method and use the resulting wave functions consistently in the decay amplitudes. 
In the factorization approximation, we evaluate $c\to s$ and $b\to c$ transitions with pion or kaon emission, including the daughter-baryon recoil in the spatial overlap. 
The calculation isolates the respective roles of the Cabibbo-Kobayashi-Maskawa (CKM) hierarchy, partial-wave kinematics, and recoil-induced cancellations in determining the decay widths and pion-to-kaon ratios.

The remainder of this paper is organized as follows. 
In Section~\ref{S2}, we introduce the constituent quark-diquark model, develop the generalized convoluted potential for heterogeneous diquarks, and present the formalism for evaluating nonleptonic weak decays within the factorization framework. 
Numerical results for the mass spectra, wave-function overlap integrals, and decay widths are presented and discussed in Section~\ref{S3}. 
Section~\ref{S4} summarizes the results, and Appendix~\ref{App:A} details the nonrelativistic reduction of the baryonic current.
The natural units $c = \hbar = 1$ are used throughout the text.

\section{Theoretical framework \label{S2}}

In this section, we present the theoretical framework for studying the spectra and nonleptonic decays of doubly and triply heavy baryons. 
We first introduce the nonrelativistic quark-diquark model used to determine the baryon masses and spatial wave functions.
Then, the formula for calculating the two-body decay widths within the factorization approximation is presented.

\subsection{Nonrelativistic quark-diquark model\label{l1}}

The triply heavy baryon $QQQ$ contain three heavy quarks, whereas the doubly heavy baryon $QQs$ considered here contain a heavy-quark pair and a strange quark.
For the heavy quarks, it is generally considered that their relative velocities inside the hadrons are sufficiently small, making a nonrelativistic treatment well-justified.
For the strange quark, owing to its larger mass compared to the light $u$ and $d$ quarks, it can be phenomenologically treated as a constituent quark with a relatively large effective mass.
Therefore, the multiply heavy baryons discussed in this paper can be treated within a nonrelativistic quark-diquark model.
The Hamiltonian for multiply heavy baryons can be written as
\begin{equation}
	H = \sum_{i}\left(m_i+\frac{p_i^2}{2m_i}\right) - T_{CM} + \sum_{i < j} V_{ij}\left(\bm{r}_{ij}\right),
	\label{1}
\end{equation}
where $m_i$ and $p_i$ denote the effective mass and momentum of the $i$-th constituent quark, respectively.
$T_{CM}$ is the center-of-mass kinetic energy of baryon, and $V_{ij}(\bm{r}_{ij})$ is the interaction between two quarks.

As with other heavy-flavor hadronic systems containing multiple heavy quarks, the dominant interaction governing the formation of bound states in these multiply heavy baryons remains the one-gluon exchange (OGE) process~\cite{DeRujula:1975qlm,Guo:2021mja,Guo:2022crh}.
To ensure precision, we employ a BCN-type interaction based on an improved OGE potential, which consists of a Cornell-like central potential and a Yukawa-smeared spin-spin term~\cite{Bhaduri:1981pn,Theussl:2000sj}.
With the color factors absorbed into the parameters, the specific expression for the interaction between two quarks reads
\begin{equation}
	V_{ij}\left(r_{ij}\right) = V_0 + C_0 r_{ij} - \frac{2\alpha_s}{3r_{ij}} + \frac{4\alpha_s\Lambda^2}{9m_i m_j}\frac{e^{-\Lambda r_{ij}}}{r_{ij}}\,\mathbf{s}_i\cdot\mathbf{s}_j,
	\label{3}
\end{equation}
where $r_{ij}=|\bm{r}_{ij}|$ and $\mathbf{s}_i$ ($\mathbf{s}_j$) is the spin operator of the $i$ ($j$)-th constituent quark. 
The symbols $V_0$, $C_0$, and $\Lambda$ are model parameters obtained by fitting to conventional hadrons.
$\alpha_s$ is a scale-dependent strong coupling constant, and its expression is 
\begin{equation}
	\alpha_s(\mu) = \frac{\alpha_0}{\ln\left(\frac{\mu^2 + \mu_0^2}{\Lambda_0^2}\right)},
	\label{p1}
\end{equation}
where the parameter $\mu$ is the reduced mass of the two quarks.
In addition, the other model parameters are consistent with those in Ref.~\cite{Liu:2025jyn}. 
They are as follows: $\Lambda_0 = 0.113$ fm$^{-1}$, $\alpha_0 = 2.118$, and $\mu_0 = 36.976 $ MeV.
Since only the $S$-wave ground states are considered, the tensor and spin-orbit terms make no contribution to the first-order interaction and are omitted.

For a diquark system, the Hamiltonian in the center-of-mass frame is written as
\begin{equation}
	H_D = m_1 + m_2 + \frac{p_r^2}{2\mu_D} + V_{12}(r),
	\label{2}
\end{equation}
where $\mu_D$ is the reduced mass of the two quarks. And $r$ is the distance between two quarks.
Solving the equation $H_D\Psi_D = M_D^\prime\Psi_D$ allows for the determination of the diquark mass $M_D^\prime$ and normalized internal wave function $\Psi_D$ for a diquark with specified quantum numbers.
Next, based on the obtained diquark mass, the Hamiltonian for the baryonic system can be constructed as
\begin{equation}
	H_B=M_D^\prime + m_3 + \frac{p_R^2}{2\mu_B}+V_{Dq}(R),
\end{equation}
where $\mu_B$ is the reduced mass of the diquark and the third quark.
$R$ is the distance of the third quark from the diquark center of mass, and $V_{Dq}(R)$ is the interaction potential between the diquark and the third quark.

The two-quark color product is $\mathbf{3}\otimes\mathbf{3}=\bar{\mathbf{3}}\oplus\mathbf{6}$. 
Only a color-antitriplet diquark can combine with the third quark to form a singlet, since $\bar{\mathbf{3}}\otimes\mathbf{3}=\mathbf{1}\oplus\mathbf{8}$, whereas $\mathbf{6}\otimes\mathbf{3}=\mathbf{8}\oplus\mathbf{10}$. 
A color-antitriplet diquark therefore transforms like an antiquark with respect to the third quark. 
The color factors are $-4/3$ for a color-singlet $q\bar q$ pair and $-2/3$ for an antitriplet $qq$ pair~\cite{Griffiths:2008zz,Kang:2025xqm}. 
Within the pairwise-potential approximation adopted here, this motivates doubling the $qq$ interaction when constructing the pointlike diquark-quark potential~\cite{Anselmino:1992vg}
\begin{equation}
	V_{Dq}^\text{unconvoluted}=2V_{qq}.
	\label{5}
\end{equation}
For an unequal-mass diquark ($m_1\neq m_2$), Eq.~\eqref{5} is insufficient because the spin-spin interaction depends explicitly on the constituent masses. 
Even in the pointlike limit, the two pairwise contributions must therefore be retained separately
\begin{equation}
	V_{Dq}^\text{unconvoluted}=V_{13}+V_{23}.
	\label{6}
\end{equation}
Here $V_{13}$ and $V_{23}$ have the form of Eq.~\eqref{3}, evaluated with the corresponding effective masses and spin operators. 
This treatment allows the pointlike approximation to neglect the finite spatial extent of the composite diquark.
To properly account for the spatial contribution of the diquark, we incorporate this effect by folding the pairwise interaction with the constituent density measured from the diquark center of mass.
For an equal-mass diquark, this construction can be written as
\begin{equation}
	V_{Dq}\left(\boldsymbol{R}\right) = \int \sigma\left(\boldsymbol{y}\right) V_{q\bar{q}}\left(|\boldsymbol{y} + \boldsymbol{R}|\right) d^3 \boldsymbol{y},
	\label{7}
\end{equation}
where $\boldsymbol{R}$ is the distance of the third quark from the diquark center of mass, and $\sigma\left(\bm{y}\right)=8|\Psi_D\left(2y\right)|^2$. 
For an $S$-wave diquark, the two-body convolution potential containing angular integrals is expressed as
\begin{equation}
	V_{Dq}\left(R\right) = \frac{1}{2} \int_0^\infty R_{nl}^2\left(r\right) I_V \left( \frac{r}{2}, R \right) r^2 dr,
	\label{8}
\end{equation}
where $R_{nl}\left(r\right)$ is the normalized diquark radial wave function.
The angular integration function can be expressed as
\begin{equation}
	I_V\left(r, R\right) = \int_{-1}^1 V_{q\bar{q}}\left(\sqrt{R^2 + r^2 + 2Rr\xi}\right) d\xi,
	\label{9}
\end{equation}
where the parameter $\xi = \cos \theta$, and $\theta$ is the angle between the two vectors $\boldsymbol{R}$ and $\boldsymbol{r}$.

\begin{figure}[htbp]
	\centering
	\includegraphics[width=0.66\columnwidth]{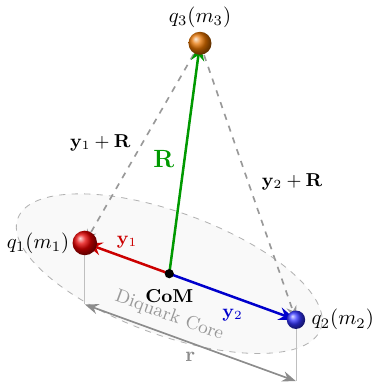}
	\caption{Coordinates used in the quark-diquark description. The relative coordinate of the diquark constituents is $\bm r$, with displacements $\bm y_1=\mu_1\bm r$ and $\bm y_2=-\mu_2\bm r$ from the diquark center of mass. $\bm R$ is defined as the vector from the center of mass of the diquark to the third constituent quark.}
	\label{fig:coordinates}
\end{figure}

For a diquark composed of two quarks with unequal masses (i.e., $m_1\neq m_2$), the displacement of the center of mass is mass-weighted, and the spin-spin interaction coefficients for the two pairs differ.
We therefore define the single-pair angular integral
\begin{equation}
	I_V^{\text{single}}\left(x,R\right)=\int_{-1}^1V_{q_iq_j}\left(\sqrt{R^2+x^2+2Rx\xi^\prime}\right)d\xi^\prime.
	\label{10}
\end{equation}
The generalized convolution potential is written as
\begin{equation}
	V_{Dq}\left(R\right) = \frac{1}{2} \int_0^\infty R_{nl}^2\left(r\right) \left(I_{V1}^{\text{single}}+I_{V2}^{\text{single}}\right) r^2 dr,
	\label{11}
\end{equation}
where the dimensionless weights are $\mu_1=m_2/\left(m_1+m_2\right)$ and $\mu_2=m_1/\left(m_1+m_2\right)$, respectively.
The symbol $I_{V1}^{\text{single}}=I_V^{\text{single}} \left( \mu_1r, R \right)$ and $I_{V2}^{\text{single}}=I_V^{\text{single}} \left( \mu_2r, R \right)$.
The corresponding coordinates are shown in Fig.~\ref{fig:coordinates}. 
Clearly, Eq.~\eqref{11} preserves both the finite size and mass asymmetry of the diquark, as well as the pairwise interactions between the quarks.
Without loss of generality, for a diquark with equal constituent quark masses, and considering $\mu_1 = \mu_2 = 1/2$, $2\, V_{qq} = V_{q\bar{q}}$ as well as the consistency of the spin-spin interaction, the diquark-quark convolution potential Eq.~\eqref{11} reduces to Eq.~\eqref{8}.

Having now obtained the arbitrary quark-diquark interaction potential, we substitute it into the Hamiltonian \eqref{1} to determine the mass spectra and wave functions of the ground-state multiply heavy baryons.

\subsection{Two-body factorized nonleptonic decay width  \label{l2}}

Using the baryon masses and wave functions obtained above, we now consider the two-body nonleptonic decays $B_i\to B_f M$, where $M$ denotes a light pseudoscalar meson. 
Of course, the decay process of a heavy quark is inevitably accompanied by the production of a virtual $W$ boson.
At the quark level, these processes are induced by the transition $Q \to q^\prime +q_1 \bar{q}_2$, with $Q=b,c$ and $q^\prime=c,s$. 
Integrating out the $W$ boson at the electroweak scale generates local four‑fermion operators with the standard vector‑minus‑axial‑vector (V-A) charged current structure~\cite{Buchalla:1995vs}. 
Short-distance QCD effects associated with the renormalization-group evolution are incorporated into the corresponding Wilson coefficients. 
By means of the operator product expansion, the effective weak Hamiltonian reads
\begin{equation}
	\mathcal{H}_{\rm eff} = \frac{G_F}{\sqrt{2}} V_{\rm CKM}\left(C_1 O_1 + C_2 O_2\right) + \mathrm{H.c.},
	\label{13}
\end{equation}
where $G_F$ is the Fermi constant, governing the universal strength of the weak interaction.
$V_{\mathrm{CKM}}\equiv V_{Q q^\prime}V_{q_1 q_2}^{*}$ denotes the product of CKM matrix elements associated with the heavy quark transition and the formation of the emitted meson.
For otherwise comparable channels, the hierarchy $\lvert V_{bc}\rvert\simeq 0.042 \ll \lvert V_{cs}\rvert\simeq 0.97$ suppresses the $b\to c$ partial widths relative to the Cabibbo-favored $c\to s$ modes~\cite{ParticleDataGroup:2024cfk}. 
Although it does not directly determine the branching fractions of parent-hadron decays, it reflects, to a certain extent, the fact that the branching fractions of bottom hadron decays are generally much smaller than those of charm hadron decays.

The symbols $C_1$ and $C_2$ are the Wilson coefficients that incorporate the effects of QCD renormalization group evolution.
Specifically, the short-distance QCD corrections arising between the electroweak scale and the strong characteristic scale are encoded in the Wilson coefficients $C_1$ and $C_2$.
For the color-favored external $W$ boson emission, these coefficients enter the factorized amplitude through an effective color‑flavor factor
\begin{equation}
a_1=C_1+\frac{C_2}{N_c},
\end{equation}
where $N_c=3$.
In the numerical calculations, we use the phenomenological values $a_1=1.03$ for bottom hadron decays and $a_1=1.10$ for charm hadron decays~\cite{Buras:1998raa}. 
These values should be regarded as effective inputs that incorporate the leading short-distance QCD effects within the factorized treatment.

The four‑fermion operators are expressed as
\begin{equation}
	\begin{split}
		O_1 = \left(\bar{q}_{1\alpha} q_{2\beta}\right)_{V-A} \left(\bar{q}^\prime_\beta Q_\alpha\right)_{V-A}, \\
		O_2 = \left(\bar{q}_{1\alpha} q_{2\alpha}\right)_{V-A} \left(\bar{q}^\prime_\beta Q_\beta\right)_{V-A},
	\end{split}
	\label{14}
\end{equation}
with
\begin{equation}
	\left(\bar{q}_1 q_2\right)_{V-A} = \bar{q}_1 \gamma_\mu P_L^\prime q_2.
	\label{15}
\end{equation}
Here, we define $P_L^\prime = 1 - \gamma_5$.

The baryonic transition matrix element remains intrinsically nonperturbative. 
For the color-favored two-body decay modes considered here, the factorization approximation is motivated by the color-transparency argument: when the emitted color-singlet meson carries sufficiently large energy, its soft-gluon interactions with the recoiling baryon are expected to be suppressed~\cite{Bjorken:1988kk}. 
Neglecting nonfactorizable soft-gluon exchanges between the two final-state hadrons, the three-hadron matrix element can be approximated by the product of a vacuum-to-meson current and a baryon transition current.
Thus, the matrix element is constructed as
\begin{equation}
	\begin{split}
		& \langle B_f M | \mathcal{H}_{\rm eff} | B_i \rangle \approx \frac{G_F}{\sqrt{2}} V_{\rm CKM} a_1 \\
		& \quad \times \langle M | \bar{q}_1 \gamma^\mu P_L^\prime q_2 |0\rangle \langle B_f | \bar{q}^\prime \gamma_\mu P_L^\prime Q | B_i \rangle.
	\end{split}
	\label{16}
\end{equation}

For a pseudoscalar meson, parity conservation implies that the vacuum-to-meson vector current matrix element vanishes. 
With the phase convention used here, the axial current matrix element is parameterized by
\begin{equation}
	\langle M\left(p\right) | \bar{q}_1 \gamma^\mu P_L^\prime q_2 | 0 \rangle = i f_M p^\mu,
	\label{17}
\end{equation}
where the symbols $f_M$ and $p^{\mu}$ denote the meson's decay constant and four-momentum, respectively.

The remaining baryon matrix element is evaluated in the nonrelativistic constituent quark model. 
At leading order, the baryon state factorizes into spin-flavor and spatial components, and the squared amplitude can be expressed in terms of the spin-flavor factors $\mathcal{C}_{V\left( A \right)}^2$ and a spatial overlap integral $I$. 
The reduction and the connection to heavy-quark spin symmetry are given in Appendix~\ref{App:A}.
In the spectator convention~\cite{Perez-Marcial:1989sch}, the daughter-baryon recoil introduces the phase factor $e^{i\bm \kappa\cdot\bm r}$, with
$\bm \kappa=\left(m_{\rm sp}/m_f\right)\bm k$. 
Here $\bm k$ is the meson momentum, $m_{\rm sp}$ is the spectator diquark mass, and $m_f$ is the sum of the final active quark and spectator masses used in the nonrelativistic coordinate transformation. 
For $S$-wave baryons, the angular integration naturally gives a zeroth-order spherical Bessel function, $j_0\left(\kappa r\right) = \sin\left(\kappa r\right)/\left(\kappa r\right)$.
The overlap integral is thus simplified to 
\begin{equation}
	I=\int_0^{\infty} R_f\left(r\right)R_i\left(r\right)j_0\left(\kappa r\right)r^2\,dr,
	\label{overlap_integral}
\end{equation}
where $R_i$ and $R_f$ are normalized radial wave functions obtained from the quark-diquark Schr\"odinger equations. 
Increasing $\kappa$ enhances cancellations in the integral and can suppress the transition amplitude.

We combine the factorized decay amplitude with the leading nonrelativistic reduction of the baryonic current. 
After integrating over the two-body phase space, the partial width for $B_i\to B_f M$ is given by
\begin{equation}
	\Gamma=\frac{G_F^2}{4\pi}|V_{\rm CKM} a_1 f_M|^2
	\frac{|\bm{k}|M_f}{M_i}
	\varepsilon I^2,
	\label{final_width}
\end{equation}
with
\begin{equation}
     \varepsilon = \left( m_M^2+|\bm{k}|^2 \right) \mathcal{C}_V^2+|\bm{k}|^2\mathcal{C}_A^2.
\end{equation}
Here, $M_i$ ($M_f$) is the initial (final) baryon mass, and $m_M$ is the final meson mass. 
Eq.~\eqref{final_width} is the working expression used below. 
Its accuracy is limited by the nonrelativistic current reduction, the spectator convention, and factorization approximation. 
These limitations should be borne in mind when interpreting the absolute widths.

\section{Results and discussion\label{S3}}

Having established the theoretical framework, we now present numerical results for the mass spectra and two-body nonleptonic weak decays of doubly and triply heavy baryons. 
We solve the radial Schr\"odinger equations using the Numerov method and employ the resulting wave functions to evaluate the spatial overlap integrals and decay widths. 
We then calculate the pion- and kaon-emission widths for the $c\to s$ and $b\to c$ transitions and examine the effects of the CKM hierarchy, phase space, and recoil dynamics.

\subsection{Model parameters\label{input}}

To obtain the mass spectra and decay widths of doubly and triply heavy baryons, we first need to extract the model parameters required for the calculations.
By fitting experimental values for conventional hadrons, we obtained the parameters for the two-body interactions: $V_0 = -0.409\,\mathrm{GeV}$, $C_0 = 0.121\,\mathrm{GeV}^2$, and $\Lambda = 0.532\,\mathrm{GeV}$.
In addition, other model parameters required for the calculations are listed in Table~\ref{tab1}.
It should be noted that the effective masses of these constituent quarks are likewise obtained by fitting to experimental observations of conventional hadrons.
The Fermi coupling constant, CKM matrix elements, and properties of light mesons are taken from the Particle Data Group~\cite{ParticleDataGroup:2024cfk}.

\begin{table}[htbp]
	\caption{Model parameters used for numerical calculations. }
	\label{tab1}
	\begin{ruledtabular}
		\renewcommand{\arraystretch}{1.3}
		\begin{tabular}{cccc} 
			Parameter & Value & Parameter & Value \\
			\hline
			$G_F$ & $1.1664 \times 10^{-5} \text{ GeV}^{-2}$ & $V_{ud}$ & $0.97435$ \\
			$m_b$ & $5.0290\text{ GeV}$ & $V_{us}$ & $0.22501$ \\
			$m_c$ & $1.7117\text{ GeV}$ & $V_{cs}$ & $0.97349$ \\
			$m_s$ & $0.4865\text{ GeV}$ & $V_{cb}$ & $0.04183$ \\
			$m_\pi$ & $0.1396 \text{ GeV}$ & $f_\pi$ & $0.130 \text{ GeV}$ \\
			$m_K$ & $0.4937 \text{ GeV}$ & $f_K$ & $0.156 \text{ GeV}$ \\ 
		\end{tabular}
	\end{ruledtabular}
\end{table}

As shown in Sec.~\ref{S2}, the nonrelativistic transition amplitudes factor into spatial overlap integrals and the spin-flavor coefficients. 
The vector and axial coefficients are denoted by $\mathcal{C}_V$ and $\mathcal{C}_A$, respectively.
Their squared values depend on the spin of the final baryon and on the number $N$ of identical quarks in the initial baryon that can undergo the weak transition. 
For ground-state triply heavy baryons, the symmetry of the spin-flavor wave functions produces the corresponding combinatorial enhancement. 
Table~\ref{tab2} lists these coefficients for all transitions considered. 
Together with the parameters in Table~\ref{tab1} and the calculated baryon masses and wave functions, they determine the decay widths.

\begin{table}[htbp]
	\caption{Values of the algebraic spin-flavor factors $\mathcal{C}_{V\left( A \right)}^2$.}
	\label{tab2}
	\centering
	\renewcommand{\arraystretch}{1.8} 
	\setlength{\tabcolsep}{5pt}
	\begin{ruledtabular}
		\begin{tabular}{c c c c c}
			$N$ & Example channels & Transition & $\mathcal{C}_V^2$ & $\mathcal{C}_A^2$ \\
			\hline
			\multirow{2}{*}{$3$} 
			& \multirow{2}{*}{\begin{tabular}{@{}c@{}} $\Omega_{bbb}^- \to \Omega_{bbc}^0$ \\ $\Omega_{ccc}^{++} \to \Omega_{cc}^+$ \end{tabular}} 
			& $3/2^+ \to 3/2^+$ & $3$ & $\frac{5}{3}$ \\
			& & $3/2^+ \to 1/2^+$ & $0$ & $\frac{4}{3}$ \\
			\hline
			\multirow{2}{*}{$2$} 
			& \multirow{2}{*}{\begin{tabular}{@{}c@{}} $\Omega_{bcc}^+ \to \Omega_{bc}^0$ \\ $\Omega_{bbc}^0 \to \Omega_{bcc}^+$ \end{tabular}} 
			& $3/2^+ \to 3/2^+$ & $2$ & $\frac{10}{9}$ \\
			& & $3/2^+ \to 1/2^+$ & $0$ & $\frac{8}{9}$ \\
			\hline
			\multirow{2}{*}{$1$} 
			& \multirow{2}{*}{\begin{tabular}{@{}c@{}} $\Omega_{bbc}^0 \to \Omega_{bb}^-$ \\ $\Omega_{bcc}^+ \to \Omega_{ccc}^{++}$ \end{tabular}} 
			& $3/2^+ \to 3/2^+$ & $1$ & $\frac{5}{9}$ \\
			& & $3/2^+ \to 1/2^+$ & $0$ & $\frac{4}{9}$ \\
		\end{tabular}
	\end{ruledtabular}
\end{table}

\subsection{Wave functions and mass spectra of doubly and triply heavy baryons  \label{Mass}}

\begin{figure}[htbp]
	\centering
	\includegraphics[width=\columnwidth]{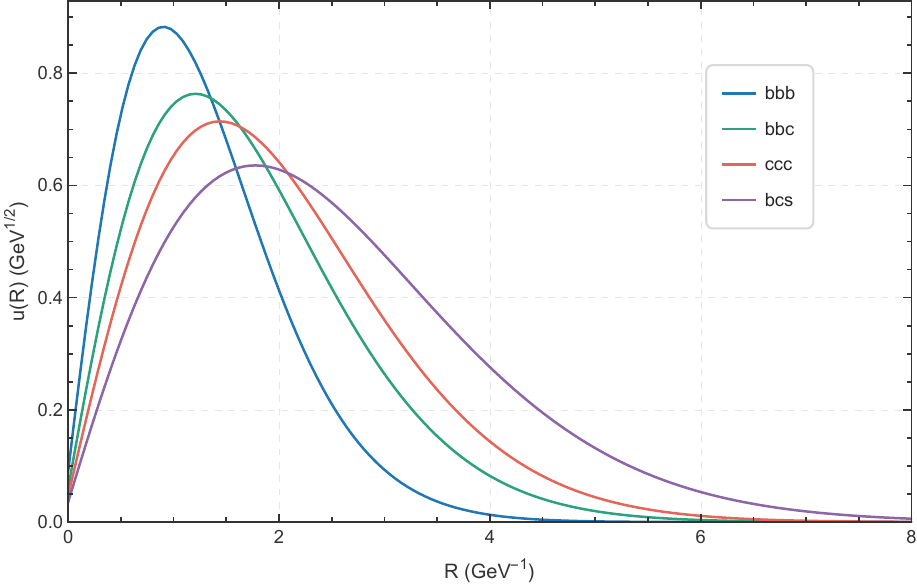}
	\caption{Reduced radial wave functions $u(r)=rR(r)$ for representative ground-state multiply heavy baryons.}
	\label{fig:wavefunctions}
\end{figure}

Using the parameters specified above, we solve the radial Schr\"odinger equations within the nonrelativistic constituent quark–diquark model. 
We obtained the radial wave functions and mass spectra of ground-state multiply heavy baryons, which were then used directly in decay calculations.
Fig.~\ref{fig:wavefunctions} shows the reduced radial wave functions $u(r)=rR(r)$ for several representative ground states. 
The corresponding mass spectra of doubly and triply heavy baryons are listed in Table~\ref{tab3}, along with results from selected lattice QCD and other models.

The wave functions in Fig.~\ref{fig:wavefunctions} satisfy the boundary conditions $u(0)=0$ and $u(r\to\infty)=0$. 
Each wave function has a single maximum, as expected for a nodeless ground state. 
The distributions broaden as the effective reduced mass decreases. 
Their maxima shift from $r\simeq0.92~\mathrm{GeV}^{-1}$ for the $bbb$ system to $r\simeq1.78~\mathrm{GeV}^{-1}$ for the $bcs$ system. 
Over the same sequence, the peak value decreases from approximately $0.883$ to $0.636~\mathrm{GeV}^{1/2}$. 
This behavior follows from the normalization condition $\int_0^\infty |u(r)|^2\,dr=1$: a broader distribution has a lower maximum. 
Among the states shown, the $bbb$ system is the most compact, whereas the $bcs$ system is the most spatially extended. 
The latter has a visible tail beyond $r\simeq8~\mathrm{GeV}^{-1}$, reflecting the presence of the lighter strange quark.

\begin{table*}[htbp]
	\caption{Predicted masses (in GeV) of ground-state doubly and triply heavy baryons, compared with other theoretical models and lattice QCD.}
	\label{tab3}
	\begin{ruledtabular}
		\small
		\renewcommand{\arraystretch}{1.3}
		\setlength{\tabcolsep}{3pt}
		\begin{tabular}{lcccccccc}
			Baryon & Components & $J^P$ & This work & \makecell{Zhao \textit{et al}. \\ \cite{Zhao:2022vfr}} & \makecell{Roberts \textit{et al}. \\ \cite{Roberts:2007ni}} & \makecell{Faustov \& \\ Ebert \textit{et al}. \cite{Faustov:2021qqf,Ebert:2002ig}} & \makecell{Ortiz-Pacheco \\ \textit{et al}. \cite{Ortiz-Pacheco:2023kjn}} & Lattice QCD \\
			\hline
			$\Omega_{bbb}^-$    & $bbb$ & $3/2^+$ & 14.336& 14.43 & --- & 14.468 & 14.861 ± 0.005 & 14.371 \cite{Meinel:2010pw} \\
			\hline
			$\Omega_{bbc}^0$    & \multirow{2}{*}{$bbc$} & $1/2^+$ & 11.165 & 11.22 & --- & 11.198 & --- & 11.195 \cite{Brown:2014ena} \\
			$\Omega_{bbc}^{*0}$ &  & $3/2^+$ & 11.170 & 11.23 & --- & 11.217 & --- & 11.229 \cite{Brown:2014ena} \\
			\hline
			$\Omega_{bcc}^+$    & \multirow{2}{*}{$bcc$} & $1/2^+$ & 8.005  & 8.02 & --- & 7.984  & --- & 8.007 \cite{Brown:2014ena} \\
			$\Omega_{bcc}^{*+}$ &  & $3/2^+$ & 8.012  & 8.03 & --- & 7.999  & --- & 8.037 \cite{Brown:2014ena} \\
			\hline
			$\Omega_{ccc}^{++}$ & $ccc$ & $3/2^+$ & 4.816  & 4.81 & --- & 4.712  & 4.903 ± 0.003 & 4.810 \cite{Padmanath:2013zfa} \\
			\hline
			$\Omega_{bb}^-$     & \multirow{2}{*}{$bbs$} & $1/2^+$ & 10.148 & --- & 10.454 & 10.359 & 10.438 ± 0.003 & 10.273 \cite{Brown:2014ena} \\
			$\Omega_{bb}^{*-}$  &  & $3/2^+$ & 10.159 & --- & 10.486 & 10.389 & 10.460 ± 0.004 & 10.308 \cite{Brown:2014ena} \\
			\hline
			$\Omega_{bc}^0$     & \multirow{2}{*}{$bcs$} & $1/2^+$ & 6.970  & --- & 7.136  & 7.088  & ---  & 6.998 \cite{Brown:2014ena} \\
			$\Omega_{bc}^{*0}$  &  & $3/2^+$ & 6.989  & --- & 7.187  & 7.130  & ---  & 7.059 \cite{Brown:2014ena} \\
			\hline
			$\Omega_{cc}^+$     & \multirow{2}{*}{$ccs$} & $1/2^+$ & 3.746  & --- & 3.815  & 3.778  & 3.766 ± 0.002  & 3.712 \cite{Mathur:2018rwu} \\
			$\Omega_{cc}^{*+}$  &  & $3/2^+$ & 3.773  & --- & 3.876  & 3.872  &  3.833 ± 0.002  & 3.788 \cite{Mathur:2018rwu} \\
		\end{tabular}
	\end{ruledtabular}
\end{table*}

In Table~\ref{tab3}, it can be seen that our mass predictions for the multiply heavy baryons are generally consistent with other theoretical models and existing lattice QCD results.
Notably, the calculated mass for $\Omega_{cc}^+$ aligns with recent experimental observations, with a discrepancy of only $0.5\%$.
Compared to lattice QCD results, the absolute differences range from $2$ to $149$~MeV, while the relative differences remain below $1.5\%$.
For the fully heavy systems, the differences do not exceed $59$~MeV, or approximately $0.6\%$.
Larger deviations occur in some strange-containing systems. 
The largest is $149$~MeV for the $\Omega_{bb}^{*-}$ baryon. The differences have no uniform sign. 
Thus, the present calculation does not show a systematic upward or downward shift relative to lattice QCD. 
The spread among phenomenological predictions is larger. 
For example, our $\Omega_{bbb}^{-}$ mass differs from the result of Ref.~\cite{Ortiz-Pacheco:2023kjn} by $0.525$~GeV, or about $3.5\%$. 
These comparisons indicate that the spin-independent part of the model provides a reasonable description of the overall mass scale. 
The remaining differences reflect the model dependence of the constituent masses, confinement potential, and relativistic corrections.

The spin-dependent interaction can be examined through the hyperfine splitting
\begin{equation}
	\Delta M=M(3/2^+)-M(1/2^+).
\end{equation}
Using the masses displayed in Table~\ref{tab3}, we obtain
\begin{equation}
	\begin{split}
		\Delta M_{bbc}&=5~\mathrm{MeV}, \qquad
		\Delta M_{bcc}=7~\mathrm{MeV},\\
		\Delta M_{bbs}&=11~\mathrm{MeV}, \qquad
		\Delta M_{bcs}=19~\mathrm{MeV},\\
		\Delta M_{ccs}&=27~\mathrm{MeV}.
	\end{split}
\end{equation}

The splittings increase from the $bbs$ to the $bcs$ and $ccs$ systems. 
This trend is consistent with the inverse-mass dependence of the chromomagnetic interaction. 
Their magnitudes, however, are systematically smaller than the corresponding lattice QCD values of $35$, $61$, and $76$~MeV~\cite{Brown:2014ena,Mathur:2018rwu}. 
The $bbc$ and $bcc$ splittings are also smaller than the lattice QCD values of $34$ and $30$~MeV \cite{Brown:2014ena}. 
Thus, the agreement obtained for the absolute masses does not extend to the hyperfine splitting. 
This discrepancy suggests that the effective spin-spin interaction is too weak in the present parametrization. 
It may also reflect the finite-range regularization and the short-distance behavior of the calculated wave functions.

\begin{table}[htbp]
	\centering
	\caption{Comparison of the ground-state masses (in GeV) of multiply heavy baryons calculated using different quark-diquark clustering configurations.}
	\label{tab:config_comparison}
	\begin{ruledtabular}
		\begin{tabular}{lccc}
			System & Configuration & $J^P = 1/2^+$ & $J^P = 3/2^+$ \\
			\hline
			$bbc$ & $(bb)c$  & 11.165 & 11.170 \\
			& $(bc)b$           & 11.191 & 11.194 \\
			\hline
			$bcc$ & $(bc)c$  &  8.005 &  8.012 \\
			& $(cc)b$           &  8.020 &  8.024 \\
			\hline
			$bbs$ & $(bb)s$  & 10.148 & 10.159 \\
			& $(bs)b$           & 10.210 & 10.214 \\
			\hline
			$bcs$ & $(bc)s$  &  6.970 &  6.989 \\
			& $(bs)c$           &  7.014 &  7.026 \\
			& $(cs)b$           &  7.021 &  7.027 \\
			\hline
			$ccs$ & $(cc)s$  &  3.746 &  3.773 \\
			& $(cs)c$           &  3.790 &  3.805 \\
		\end{tabular}
	\end{ruledtabular}
\end{table}

It should be noted that for the five mixed-flavor baryonic systems $bbc$, $bcc$, $bbs$, $bcs$, and $ccs$, we group the two heaviest quarks into a color-antitriplet diquark and adopt the quark-diquark clustering configurations $(bb)c$, $(bc)c$, $(bb)s$, $(bc)s$, and $(cc)s$, respectively. 
This strategy is based on a simple dynamic motivation.
At a given relative momentum, a pair with a larger reduced mass has a lower relative kinetic energy and can therefore localize at shorter distances at a smaller kinetic cost. 
The mass spectra of multiply heavy baryons also depend on the interaction of the finite-size diquark with the third quark. 
Naturally, we also calculated the ground-state masses of multiply heavy baryons for various quark-diquark clustering configurations, as shown in Table~\ref{tab:config_comparison}.
For all five systems, the heaviest-pair configuration gives the lowest calculated mass in both the quantum numbers $J^P=1/2^+$ and $3/2^+$. 
This ordering supports our choice within the quark-diquark approximation.

For the $bcc$ baryon, exchanging the two charm quarks maps $(bc_1)c_2$ onto $(bc_2)c_1$.
The color wave function is antisymmetric under this exchange and the $cc$ flavor state is symmetric, so the combined spin and spatial wave function must be symmetric. 
In the $(cc)b$ configuration, the internal $S$-wave diquark $cc$ is spatially symmetric and therefore requires $S_{cc}=1$. 
In a single $(bc)c$ configuration, imposing $l_D=l_{Dq}=0$ in the chosen coordinates does not enforce charm-quark exchange symmetry. 
In particular, any $S_{cc}=0$ component of the physical $J^P=1/2^+$ state must be accompanied by a spatial component that is antisymmetric under $c_1\leftrightarrow c_2$.
Of course, the present calculation does not quantify the effect of enforcing this condition. 
In Table~\ref{tab:config_comparison}, relative to $(cc)b$ configuration, the masses of $(bc)c$ are lower by 15 MeV for $J^P=1/2^+$ and 12 MeV for $J^P=3/2^+$. 
These differences measure the sensitivity to the clustering choice. 
Assessing the effect of exchange symmetry on the masses would require combining the two exchange-related $(bc)c$ configurations with the required symmetry or performing an explicitly antisymmetrized three-body calculation.

\subsection{Spatial overlap integrals \label{SOI}}

\begin{table}[htbp]
	\caption{Meson momenta $k$ (in GeV) and dimensionless spatial overlap integrals $I$ for the two-body weak decays considered in this work. 
		All initial baryons have $J_i^P=3/2^+$.}
	\label{tab4}
	\begin{ruledtabular}
		\small
		\renewcommand{\arraystretch}{1.3}
		\setlength{\tabcolsep}{3pt}
		\begin{tabular}{llccc}
			Transition & Decay channel & $J_i^P \to J_f^P$ & $k$  &  $I$ \\
			\hline
			\multicolumn{5}{l}{$b \to c$ Nonleptonic decays} \\
			\quad $N=3$ & $\Omega_{bbb}^- \to \Omega_{bbc}^0 \pi^-$    & $3/2^+ \to 1/2^+$ & 2.8171 & 0.1882 \\
			& $\Omega_{bbb}^- \to \Omega_{bbc}^{*0} \pi^-$  & $3/2^+ \to 3/2^+$ & 2.8135 & 0.1876 \\
			& $\Omega_{bbb}^- \to \Omega_{bbc}^0 K^-$      & $3/2^+ \to 1/2^+$ & 2.7849 & 0.1942 \\
			& $\Omega_{bbb}^- \to \Omega_{bbc}^{*0} K^-$    & $3/2^+ \to 3/2^+$ & 2.7812 & 0.1936 \\
			\cline{2-5} 
			\quad $N=2$ & $\Omega_{bbc}^{*0} \to \Omega_{bcc}^+ \pi^-$  & $3/2^+ \to 1/2^+$ & 2.7136  & 0.1519 \\
			& $\Omega_{bbc}^{*0} \to \Omega_{bcc}^{*+} \pi^-$ & $3/2^+ \to 3/2^+$ & 2.7084  & 0.1512 \\
			\cline{2-5}
			\quad $N=1$ & $\Omega_{bcc}^{*+} \to \Omega_{ccc}^{++} \pi^-$ & $3/2^+ \to 3/2^+$ & 2.5560  & 0.2340 \\
			\hline
			\multicolumn{5}{l}{$c \to s$ Nonleptonic decays} \\
			\quad $N=3$ & $\Omega_{ccc}^{++} \to \Omega_{cc}^+ \pi^+$   & $3/2^+ \to 1/2^+$ & 0.9433  & 0.5854 \\
			& $\Omega_{ccc}^{++} \to \Omega_{cc}^{*+} \pi^+$ & $3/2^+ \to 3/2^+$ & 0.9217  & 0.5899 \\
			& $\Omega_{ccc}^{++} \to \Omega_{cc}^+ K^+$     & $3/2^+ \to 1/2^+$ & 0.8429  & 0.6458 \\
			& $\Omega_{ccc}^{++} \to \Omega_{cc}^{*+} K^+$   & $3/2^+ \to 3/2^+$ & 0.8181  & 0.6525 \\
			\cline{2-5}
			\quad $N=2$ & $\Omega_{bcc}^{*+} \to \Omega_{bc}^0 \pi^+$    & $3/2^+ \to 1/2^+$ & 0.9655  & 0.5672 \\
			& $\Omega_{bcc}^{*+} \to \Omega_{bc}^{*0} \pi^+$  & $3/2^+ \to 3/2^+$ & 0.9490  & 0.5707 \\
			\cline{2-5}
			\quad $N=1$ & $\Omega_{bbc}^{*0} \to \Omega_{bb}^- \pi^+$    & $3/2^+ \to 1/2^+$ & 0.9661  & 0.5773 \\
			& $\Omega_{bbc}^{*0} \to \Omega_{bb}^{*-} \pi^+$  & $3/2^+ \to 3/2^+$ & 0.9553  & 0.5791 \\
		\end{tabular}
	\end{ruledtabular}
\end{table}

Table~\ref{tab4} lists the meson momentum $k$ and the spatial overlap integral $I$ for each decay channel. 
The overlap integral is defined in Eq.~\eqref{overlap_integral}.
Its recoil dependence enters through the argument of the spherical Bessel function, $j_0(\kappa r)$, where $\kappa=(m_{\rm sp}/m_f)k$. 
Thus, $k$ determines the physical two-body phase space, whereas $\kappa$ is the momentum that enters the internal quark-diquark coordinate. 
$N$ denotes the number of identical quarks in the initial baryon that can undergo the weak transition.

The clearest pattern in Table~\ref{tab4} is the separation between the two quark-level transitions. 
The $b\to c$ overlaps range from $0.1512$ to $0.2340$, whereas the $c\to s$ values range from $0.5672$ to $0.6525$. 
The corresponding meson momenta are $2.5560$ -- $2.8171$~GeV and $0.8181$ -- $0.9661$~GeV, respectively. 
Because $\kappa$ is proportional to $k$, the larger recoil in the $b\to c$ channels produces more rapid oscillations of $j_0(\kappa r)$. 
This enhances cancellations in the radial integral and suppresses the spatial overlap integral $I$. 
The separation between the two sectors is therefore consistent with recoil suppression. 
The radial wave functions also contribute, however, because different initial and final baryons are involved. 
Indeed, the nonmonotonic variation of $I$ within the $b\to c$ sector shows that the overlap is not determined by $k$ alone.

The comparison of pion and kaon emission provides a more direct test of the recoil dependence because the baryonic wave functions are unchanged. 
For a fixed baryonic transition, the larger kaon mass reduces the available kinetic energy and hence the momentum $k$. 
The weaker oscillation of $j_0(\kappa r)$ then increases the overlap. 
For example, for fully heavy baryon $\Omega_{bbb}^-\to\Omega_{bbc}^0 M^-$, the overlap $I$ increases from $0.1882$ for $M=\pi$ to $0.1942$ for $M=K$. 
For $\Omega_{ccc}^{++}\to\Omega_{cc}^+M^+$, it increases from $0.5854$ to $0.6458$.
The larger relative change in the $c\to s$ channel reflects the greater fractional reduction of $k$ when the pion is replaced by a kaon.

\subsection{Nonleptonic decay widths \label{subsec:decay_widths}}

\begin{figure}[htbp]
	\centering
	\includegraphics[width=\columnwidth]{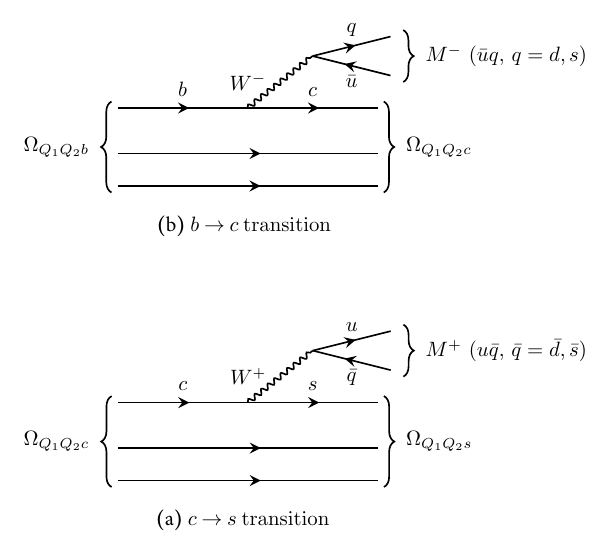}
	\caption{External $W$ boson emission topologies for the nonleptonic transitions considered in this work: (a) $c\to s$ with $M^+ = u\bar q $ and
		(b) $b\to c$ with $M^- = \bar u q $, where $q=d,s$.}
	\label{fig:weak-decay}
\end{figure}

We now combine the kinematic factors, spin-flavor coefficients, and spatial overlap integrals in Eq.~\eqref{final_width}. 
The external $W$ emission topologies are shown in Fig.~\ref{fig:weak-decay}. 
In Fig.~\ref{fig:weak-decay}(a), an active charm quark undergoes $c\to sW^+$, followed by $W^+\to u\bar q$ with a light antiquark $\bar q = \bar d$ or $\bar s$. 
In Fig.~\ref{fig:weak-decay}(b), the corresponding transition is $b\to cW^-$, followed by $W^-\to q\bar u$. 
In both cases, the two remaining quarks from the initial baryon act as spectators.

The calculated $c\to s$ widths are listed in Table~\ref{tab:decay_widths_cs}. 
Clearly, our results extend to a broader range of heavy-flavor baryon decays.
The results include the CKM factors, the effective coefficient $a_1$, the meson decay constants, two-body phase space, spin-flavor coefficients, and recoil-dependent spatial overlaps.

\begin{table*}[!t]
	\caption{Calculated nonleptonic decay widths, in units of $10^{-16}$~GeV, for the $c\to s$ transitions. Available light-front quark model results are included for comparison.}
	\label{tab:decay_widths_cs}
	\begin{ruledtabular}
		\renewcommand{\arraystretch}{1.3}
		\begin{tabular}{lccccc} 
			Decay Channel & $J_i^P \to J_f^P$ & This work & LFQM~(Wang~\cite{Wang:2022}) & LFQM~(Lu~\cite{Lu:2024}) & LFQM~(Zhao~\cite{Zhao:2022vfr}) \\
			\hline
			$\Omega_{ccc}^{++} \to \Omega_{cc}^+ \pi^+$ & $3/2^+ \to 1/2^+$ & 594.030 & $732 \pm 160 \pm 457 \pm 224 \pm 1$ & 640.440/803.019 & 291\\
			$\Omega_{ccc}^{++} \to \Omega_{cc}^{*+} \pi^+$ & $3/2^+ \to 3/2^+$ & 2013.77 & --- & --- & --- \\
			$\Omega_{ccc}^{++} \to \Omega_{cc}^+ K^+$ & $3/2^+ \to 1/2^+$ & 39.6164 & $32.0 \pm 8.1 \pm 22.9 \pm 10.4 \pm 0.9$ & 36.3333/46.5356 & 26.2 \\
			$\Omega_{ccc}^{++} \to \Omega_{cc}^{*+} K^+$ & $3/2^+ \to 3/2^+$ & 160.875 & --- & --- & --- \\
			$\Omega_{bcc}^{*+} \to \Omega_{bc}^0 \pi^+$ & $3/2^+ \to 1/2^+$ & 445.858 & --- & --- & 210 \\
			$\Omega_{bcc}^{*+} \to \Omega_{bc}^{*0} \pi^+$ & $3/2^+ \to 3/2^+$ & 1525.99 & --- & --- & --- \\
			$\Omega_{bbc}^{*0} \to \Omega_{bb}^- \pi^+$ & $3/2^+ \to 1/2^+$ & 241.670 & --- & --- & 127 \\
			$\Omega_{bbc}^{*0} \to \Omega_{bb}^{*-} \pi^+$ & $3/2^+ \to 3/2^+$ & 835.364 & --- & --- & --- \\
		\end{tabular}
	\end{ruledtabular}
\end{table*}

For the benchmark channel $\Omega_{ccc}^{++}\to\Omega_{cc}^{+}\pi^+$, we obtain $\Gamma=5.9403\times10^{-14}$~GeV. 
This result lies within the broad uncertainty range of Ref.~\cite{Wang:2022}. 
It is also close to the two values reported in Ref.~\cite{Lu:2024}. 
The prediction of Ref.~\cite{Zhao:2022vfr} is smaller by approximately a factor of two, but it has the same order of magnitude. 
A similar level of consistency is found for $\Omega_{ccc}^{++}\to\Omega_{cc}^{+}K^+$. 
These comparisons provide a useful normalization check, although the present model uncertainties have not been quantified.

For otherwise identical channels, the widths with a $J_f^P=3/2^+$ sub-channel are larger than those with a $J_f^P=1/2^+$ sub-channel.
The enhancement factors in the $c\to s$ sector range from $3.39$ to $4.06$. 
The corresponding spatial overlaps differ by no more than about $1.1\%$, as shown in Table~\ref{tab4}. 
The enhancement therefore arises mainly from the spin-flavor structure. 
In particular, the vector contribution vanishes for the $3/2^+\to1/2^+$ transitions, whereas both vector and axial contributions are present for $3/2^+\to3/2^+$ transitions.

The pion-to-kaon ratios separate the flavor factors from the residual kinematic and overlap effects. 
If these residual effects are neglected, the reference ratio is
\begin{equation}
	\mathcal{R}_0=
	\left|\frac{V_{ud}}{V_{us}}\right|^2
	\left(\frac{f_\pi}{f_K}\right)^2
	\simeq 13.01.
	\label{eq:baseline-ratio}
\end{equation}

For the triply charmed baryon decay channels $\Omega_{ccc}^{++}\to\Omega_{cc}^{* +}M^+$, the values in Table~\ref{tab:decay_widths_cs} give
\begin{equation}
	\frac{\Gamma_{\Omega_{ccc}^{++}\to\Omega_{cc}^{+}\pi^+}}
	{\Gamma_{\Omega_{ccc}^{++}\to\Omega_{cc}^{+}K^+}}
	=15.00,
	\,\,\,\,
	\frac{\Gamma_{\Omega_{ccc}^{++}\to\Omega_{cc}^{*+}\pi^+}}
	{\Gamma_{\Omega_{ccc}^{++}\to\Omega_{cc}^{*+}K^+}}
	=12.52.
\end{equation}
For the final-state baryon with quantum numbers $J_f^P=1/2^+$, $\mathcal{C}_V^2=0$, and the leading width is proportional to $|\bm k|^3\mathcal{C}_A^2I^2$. 
The smaller kaon momentum thus raises the ratio above $\mathcal{R}_0$. 
For the quantum numbers $J_f^P=3/2^+$, the additional vector term is proportional to $|\bm k|E_M^2$. 
This term weakens the momentum dependence. 
The larger kaon overlap further lowers the ratio, giving a value slightly below $\mathcal{R}_0$.

Table~\ref{tab:decay_widths_bc} gives the corresponding $b\to c$ widths. 
They are smaller than the analogous $c\to s$ widths by factors of approximately $1.0\times10^2$ -- $4.7\times10^2$. 
This suppression is driven primarily by $|V_{cb}|^2\ll|V_{cs}|^2$ and by the smaller recoil-dependent overlaps in the $b\to c$ sector. 
The larger $b\to c$ phase space partially offsets these two effects.

\begin{table}[!htbp]
	\caption{ Calculated nonleptonic decay widths, in units of $10^{-16}$~GeV, for the $b\to c$ transitions.}
	\label{tab:decay_widths_bc}
	\begin{ruledtabular}
		\renewcommand{\arraystretch}{1.3}
		\begin{tabular}{lcc} 
			Decay channel & $J_i^P \to J_f^P$ & This work \\
			\hline
			$\Omega_{bbb}^- \to \Omega_{bbc}^0 \pi^-$ & $3/2^+ \to 1/2^+$ & 2.6525 \\
			$\Omega_{bbb}^- \to \Omega_{bbc}^{*0} \pi^-$ & $3/2^+ \to 3/2^+$ & 9.2061 \\
			$\Omega_{bbb}^- \to \Omega_{bbc}^0 K^-$ & $3/2^+ \to 1/2^+$ & 0.2096 \\
			$\Omega_{bbb}^- \to \Omega_{bbc}^{*0} K^-$ & $3/2^+ \to 3/2^+$ & 0.7411 \\
			$\Omega_{bbc}^{*0} \to \Omega_{bcc}^+ \pi^-$ & $3/2^+ \to 1/2^+$ & 0.9467 \\
			$\Omega_{bbc}^{*0} \to \Omega_{bcc}^{*+} \pi^-$ & $3/2^+ \to 3/2^+$ & 3.2737 \\
			$\Omega_{bcc}^{*+} \to \Omega_{ccc}^{++} \pi^-$ & $3/2^+ \to 3/2^+$ & 8.2851 \\
		\end{tabular}
	\end{ruledtabular}
\end{table}

The quantum numbers $J_f^P=3/2^+$ enhancement is also present in the $b\to c$ sector. 
The ratios of spin-partner widths range from $3.46$ to $3.54$. 
For the $\Omega_{bbb}^-\to\Omega_{bbc}^{(*)0}M^-$ channels, the pion-to-kaon ratios are

\begin{equation}
	\frac{\Gamma_{\Omega_{bbb}^-\to\Omega_{bbc}^{0}\pi^-}}
	{\Gamma_{\Omega_{bbb}^-\to\Omega_{bbc}^{0}K^-}}
	=12.66,
	\,\,\,\,
	\frac{\Gamma_{\Omega_{bbb}^-\to\Omega_{bbc}^{*0}\pi^-}}
	{\Gamma_{\Omega_{bbb}^-\to\Omega_{bbc}^{*0}K^-}}
	=12.42.
\end{equation}
Clearly, both values lie within about $5\%$ of $\mathcal{R}_0$.
The pion and kaon momenta differ by only about $1\%$ in these decay channels because the energy release is approximately $3.2$~GeV. 
Thus, the remaining deviations from $\mathcal{R}_0$ arise from phase space, meson-energy factors, and the slightly larger kaon overlaps.

These patterns provide internal checks of the flavor factors, recoil prescription, and leading partial-wave dependence. 
The absolute decay widths remain model dependent. 
The calculated decay widths still depend on the chosen model.
Their uncertainties receive contributions from the potential parameters, the nonrelativistic current reduction, the factorization approximation, and neglected configuration mixing. 
The tabulated values should therefore be regarded as benchmark estimates rather than precision predictions.

\section{Summary and outlook  \label{S4}}

We have investigated the mass spectra and two-body nonleptonic weak decays of ground-state doubly and triply heavy baryons in a common nonrelativistic quark-diquark framework. 
The central methodological development is the extension of the finite-size convoluted interaction to unequal-mass diquarks. 
The extension uses mass-weighted constituent coordinates and retains the distinct pairwise spin-spin couplings. 
The resulting Schr\"odinger equations provide the baryon masses and wave functions used in the decay calculation. 
Within the factorization approximation, the transition matrix elements separate into spin-flavor coefficients and recoil-dependent spatial overlaps. 
Therefore, this construction treats the spectroscopy and weak transitions with the same set of baryon wave functions.

The differences between our calculated masses for multiply heavy baryons and the cited lattice QCD values are less than $1.5\%$.
In particular, the predicted $\Omega_{cc}^{+}$ mass agrees with the recent LHCb Collaboration measurement at the $0.5\%$ level. 
However, the hyperfine splittings are systematically slightly smaller than the corresponding lattice QCD results. 
Thus, the present parametrization describes the overall mass scale more accurately than the hyperfine structure. 
For the weak transitions, the leading nonrelativistic spin-flavor algebra gives $\mathcal{C}_V^2=0$ for $3/2^+\to1/2^+$ channels. 
These modes receive only the leading axial contribution, whereas both vector and axial terms contribute to $3/2^+\to3/2^+$ transitions. 
The $b\to c$ spatial overlaps lie in the range $0.1512$ -- $0.2340$, compared with $0.5672$ -- $0.6525$ for $c\to s$. 
This separation is consistent with stronger recoil-induced cancellations in the $b\to c$ sector, although differences between the radial wave functions also
contribute. 
Together with the CKM hierarchy, this effect makes analogous $b\to c$ widths smaller than the $c\to s$ widths by factors of approximately $1.0\times10^2$ -- $4.7\times10^2$. 
The pion-to-kaon ratios further expose the partial-wave function dependence. 
For the representative $c\to s$ channels, the ratios are $15.00$ and $12.52$ for the quantum numbers $J_f^P=1/2^+$ and $3/2^+$, respectively. 
The corresponding $b\to c$ ratios, $12.66$ and $12.42$, remain close to the flavor only reference value $\mathcal{R}_0\simeq13.01$ because the relative pion-kaon phase-space difference is small.

The predicted mass spectra and partial decay widths provide benchmark estimates for future searches for multiply heavy baryons. 
Ratios of related pion- and kaon-emission modes may be more stable because several normalization uncertainties cancel. 
The results are nevertheless limited by the nonrelativistic treatment, spectator convention, factorization approximation, and the absence of a quantitative uncertainty analysis. 
A systematic study should vary the constituent effective masses and potential parameters and propagate the resulting changes to the wave functions and decay widths. 
Relativistic corrections, configuration mixing, and nonfactorizable contributions should also be assessed, especially for final states containing a strange quark. 
Finally, a direct comparison between finite-size and pointlike diquark calculations would isolate the effect of the generalized convolution potential. 
Comparisons of the resulting transition form factors with lattice QCD would provide an independent test of the framework.

\begin{acknowledgments}
	The authors thank Xi Xia for helpful comments and discussions. This work was supported by the Scientific Research Foundation of Chengdu University of Technology under Grant No. 10912-KYQD2025-09557.
\end{acknowledgments}

\appendix
\section{Nonrelativistic reduction of the baryonic matrix elements}
\label{App:A}

We summarize the leading nonrelativistic reduction of the baryonic current $J^\mu=\bar q^\prime\gamma^\mu P_L^\prime Q$, with $P_L^\prime=1-\gamma_5$.
Relativistic one-particle states satisfy $\langle\bm p^\prime,s^\prime|\bm p,s\rangle_{\rm rel} =2E(2\pi)^3\delta^{(3)}(\bm p^\prime-\bm p)\delta_{s^\prime s}$.
For the multiply heavy baryon systems, therefore, the corresponding matrix elements are related by~\cite{LeYaouanc1988}
\begin{equation}
	\begin{aligned}
		\langle B_f|J^\mu|B_i\rangle_{\rm rel}
		&=2\sqrt{E_iE_f}\,
		\langle B_f|J^\mu|B_i\rangle_{\rm NR} \\
		&\simeq 2\sqrt{M_iM_f}\,
		\langle B_f|J^\mu|B_i\rangle_{\rm NR}.
	\end{aligned}
	\label{eq:app_normalization}
\end{equation}
In the parent-baryon rest frame, $E_i=M_i$ and $E_f=\sqrt{M_f^2+|\bm k|^2}$.
The last step neglects relative corrections of order $|\bm k|^2/M_f^2$ to the normalization factor.

At leading order in the constituent velocities, the vector charge reduces to the unit operator, and the spatial axial current reduces to the Pauli operator. 
The spatial vector current and the axial charge start at first-order in these velocities and are omitted. 
Within the $S$-wave ansatz used here, the baryon state separates into spin-flavor and spatial components,
\begin{equation}
	|B_{i,f}\rangle_{\rm NR}
	=|\Phi_{i,f}\rangle\otimes|\Psi_{i,f}\rangle,
\end{equation}
where $|\Phi_{i,f}\rangle$ and $|\Psi_{i,f}\rangle$ represent the spin-flavor wave function and the spatial radial wave function, respectively.
We then define the corresponding spin-flavor operators as
\begin{equation}
	\mathcal O_V=\sum_{a=1}^{3}T_a,
	\qquad
	\bm{\mathcal O}_A=\sum_{a=1}^{3}\bm\sigma_a T_a,
	\label{eq:app_operators}
\end{equation}
where $T_a=(|q^\prime\rangle\langle Q|)_a$ changes the flavor of quark $a$, and $\bm\sigma_a$ acts on the spin of the same quark. 
Only constituents of heavy-flavor $Q$ contribute. 
Flavor multiplicities are included through the normalized spin-flavor states.

In the spectator convention, equivalent active-quark contributions share the spatial overlap integral $I$ defined in Eq.~\eqref{overlap_integral}. 
We use $g_{\mu\nu}=\operatorname{diag}(1,-1,-1,-1)$ and $\gamma_5=i\gamma^0\gamma^1\gamma^2\gamma^3$.
For the emitted meson momentum $p^\mu=(E_M,\bm k)$, the contracted current is
\begin{equation}
	\begin{aligned}
		p_\mu\langle B_f|J^\mu|B_i\rangle_{\rm rel}
		\simeq{}&2\sqrt{M_iM_f}\,I \\
		&\times\langle\Phi_f|
		E_M\mathcal O_V+\bm k\cdot\bm{\mathcal O}_A
		|\Phi_i\rangle.
	\end{aligned}
	\label{eq:app_contracted_current}
\end{equation}
The two terms describe the leading spin-independent and spin-dependent contributions, respectively.

After averaging over the initial spin and summing over the final spin, the coefficients entering Eq.~\eqref{final_width} are
\begin{equation}
	\begin{aligned}
		\mathcal C_V^2
		&=\frac{1}{2J_i+1}\sum_{s_i,s_f}
		\left|\langle\Phi_f(s_f)|\mathcal O_V|\Phi_i(s_i)\rangle\right|^2,
		\\
		\mathcal C_A^2
		&=\frac{1}{2J_i+1}\sum_{s_i,s_f}
		\left|\langle\Phi_f(s_f)|
		\hat{\bm k}\cdot\bm{\mathcal O}_A
		|\Phi_i(s_i)\rangle\right|^2,
	\end{aligned}
	\label{eq:app_spin_factors}
\end{equation}
where the symbols $s_i$ and $s_f$ denote spin projections, and $\hat{\bm k}=\bm k/|\bm k|$. 
Rotational invariance makes these spin sums independent of $\hat{\bm k}$ and eliminates the vector--axial interference term.
Thus, we obtain
\begin{equation}
	\begin{aligned}
		\overline{\sum}
		\left|p_\mu\langle B_f|J^\mu|B_i\rangle_{\rm rel}\right|^2
		\simeq{}&4M_iM_f|I|^2 \\
		&\times\left(E_M^2\mathcal C_V^2
		+|\bm k|^2\mathcal C_A^2\right).
	\end{aligned}
	\label{eq:app_spin_sum}
\end{equation}
Here $\overline{\sum}$ denotes the same spin average and sum. 
Combining this result with the factorized meson current and the two-body phase space gives Eq.~\eqref{final_width}.

The operator $\mathcal O_V$ is a scalar under spin rotations. 
It cannot connect the $J_i=3/2$ and $J_f=1/2$ states in the present $S$-wave basis, so $\mathcal C_V^2=0$. 
The axial operator has spin rank one and allows this transition. 
Its contraction with $\bm k$ produces the leading $P$-wave amplitude and the associated $|\bm k|^3|I|^2$ dependence of the width. 
This selection rule applies to the retained leading order current. 
Subleading currents and relativistic corrections require a separate assessment, particularly for a strange daughter quark.

\bibliography{ref}

\end{document}